# Can a pseudo-symmetry solve the cosmological constant problem?

S.R. Beane[†]

Department of Physics, Duke University
Durham, NC 27708

*sbeane@phy.duke.edu*

## ABSTRACT

A general no-go theorem dampens hope that the cosmological constant problem can be solved by a local symmetry mechanism. The possibility is considered here that this no-go theorem can be avoided by a pseudo-symmetry. A simple macroscopic effective field theory is constructed which admits an enhanced pseudo-symmetry in the absence of a cosmological term. It is pointed out that this pseudo-symmetry is an exact classical invariance of superstrings. The conjecture that this pseudo-symmetry survives in the quantum theory has several interesting consequences.

---

[†] This work was supported in part by the
U.S. Department of Energy (Grant DE-FG05-90ER40592).

The cosmological constant is very small at macroscopic scales. Since there is no evidence whatsoever of non-local phenomena in nature, the most conservative approach to the cosmological constant problem is to attempt to identify a workable symmetry mechanism in the macroscopic effective theory [1]. The source of the symmetry mechanism in the underlying theory is a separate issue. These statements lie at the heart of the effective field theory paradigm: One can always describe a particular physical phenomenon at the characteristic scale at which the phenomenon is relevant, without having to understand physics at other scales [2]. For example, we thankfully need not know where $U(1)$ electromagnetism comes from in the underlying theory in order to find its consequences useful at low-energies.

Here we take the conservative point of view; that is, we assume that the mechanism which renders the vacuum energy small is local. Very general considerations then require that there exist at least one quantum in the mili-eV range that has not yet been observed [1]. Unfortunately, this conservative viewpoint immediately runs into serious trouble. There exists a general "no-go" theorem which seems to eliminate the possibility of a solution based on a local symmetry argument [3]. Evidently, even if we assume the existence of unobserved quanta, there is no escaping an undesirable fine-tuning of parameters.

In this letter, we will investigate the possibility that this no-go theorem can be circumvented by a pseudo-symmetry. That is, we will allow coupling constants to transform as well as fields. To our knowledge, this approach has never before been considered in the literature. We will first construct a toy effective field theory based on a non-linearly realized scale invariance and demonstrate the content of the no-go theorem. We focus on scale invariance because we can think of no other potentially relevant symmetry with non-trivial action on the metric field. Consider a global scale transformation:

$$
\begin{aligned}
g_{\mu\nu} &\longrightarrow e^{-2\lambda} g_{\mu\nu} \\
\phi &\longrightarrow \phi + \lambda M \\
\psi_i &\longrightarrow e^{\lambda D_i} \psi_i.
\end{aligned}
\tag{1}
$$

General coordinate invariance, which is assumed, relates this transformation to one in which the coordinate scales, rather than the metric [4]. $\psi_i$ is a general field coordinate with scale dimension matrix $D_i$. The scalar field $\phi$ can be thought of as the Goldstone boson of broken scale invariance (a dilaton), as suggested by its transformation law. $M$ is then the scale of spontaneous symmetry breaking. (In what follows, M is scaled out for notational convenience.)

An interesting feature of this symmetry is that anomalies play a central role in its realization. This can be seen by way of the following argument [5]. A scale invariant potential will satisfy $V(\alpha \chi) = \alpha^4 V(\chi)$, where $\chi$ is a scalar field with canonical scale dimension. It follows that $V$ has a minimum at the origin, $V$ is flat, or $V$ is unbounded from below. Therefore, a scalar field in a scale invariant theory can have a non-vanishing vacuum expectation value (VEV) if either (i) the potential is flat, or (ii) the scale invariance is anomalous. A linearly transforming field is readily constructed out of $\phi$: $\chi \equiv e^{\phi}$. Since $\langle \chi \rangle \neq 0$ for finite[1] $\langle \phi \rangle$, if there is a non-trivial equilibrium point, either the dilaton is a pseudo-Goldstone boson or the dilaton potential is flat. This has a very interesting consequence: If a dilaton solves the cosmological constant problem, the cosmological constant is not identically zero, since there always exists a scale at which a massive dilaton can be integrated out. (For example, in the absence of the dilaton there would be nothing to protect the vacuum against electromagnetic and gravitational fluctuations.) On general grounds, one would expect $\rho_{\text{VAC}} \simeq (m_\phi)^4$. A rough cosmological bound [3] on $\rho_{\text{VAC}}$ then implies $m_\phi \lesssim 10^{-12} GeV$.

---

1 Here we ignore the possibility of a "runaway" minimum.

Consider an effective field theory with all the massive particles in nature, except the dilaton, integrated out. The cutoff can be taken as the electron mass. For simplicity, consider first a toy model in which all symmetry breaking terms in the potential transform like the generic conformal anomaly

$$\frac{\beta(g)}{2g} F^a_{\mu\nu} F^{\mu\nu}_a + m \gamma_m(g) \bar\Psi \Psi, \tag{2}$$

where there is an implicit sum over all gauge and matter fields. The effective matter action takes the form

$$S_m[\phi, g] = -\int d^4x \sqrt{g}\, V(\phi) = \int d^4x \sqrt{g}\, \{-e^{4\phi}[\Delta_0 + \phi \Delta_1 + e^{-\phi}\Delta_2]\} \tag{3}$$

where $V(\phi)$ is the potential, and $\Delta_0$, $\Delta_1$, and $\Delta_2$ are c-number coefficients. It is a straightforward excercise to show that the second and third terms in Eq. (3) transform like the first and second terms in Eq. (2), respectively. The Einstein action[2],

$$S_E[g] = -\frac{1}{2\kappa^2} \int d^4x \sqrt{g}\, R, \tag{4}$$

can be made scale invariant by the replacement $R \to R e^{2\phi}$, and yet there is little motivation to do so. Here it is treated as an additional symmetry breaking term[3]. The total action is then $S[\phi, g, A_\mu, \psi_i, ...] = S_E[g] + S_m[\phi, g] + ...$, where the dots include all derivative (including the dilaton kinetic term), and higher dimensional interactions of the dilaton, graviton, photon ($A_\mu$), neutrinos ($\psi_i$), and any undetected forms of matter (e.g. axions?), consistent with the assumed symmetries.

An equilibrium solution —constant in space and time— to the dilaton field equation can be found such that

$$0 = 4\Delta_0 + (4\phi_o + 1)\Delta_1 + 3\Delta_2 e^{-\phi_o}. \tag{5}$$

However, this is not the condition for a vanishing cosmological constant. The condition for a flat space solution of the Einstein equation is

$$0 = \Delta_0 + \phi_o \Delta_1 + \Delta_2 e^{-\phi_o}. \tag{6}$$

---

2 Throughout, I follow the conventions of Ref. 3.

3 For consistency one should include a term $e^{-2\phi}\Delta_3$ in (the square brackets of) the potential which transforms like the Einstein action. However, its inclusion adds no insight to the analysis.

This constitutes an example of the no-go theorem. Evidently, scale invariance, in itself, is not enough to ensure a vanishing cosmological constant [3,6,7]. Nevertheless, one might consider the significance of what is missing. Eqs. (5) and (6) imply $\Delta_1 = \Delta_2 e^{-\phi_o}$. Under an infinitesimal scale transformation, $S_m$ transforms as

$$S_m[\phi, g] \to \tilde{S}_m[\phi, g] = S_m[\phi, g] + \epsilon \int d^4x \sqrt{g} \{-e^{4\phi}[\Delta_1 - \Delta_2 e^{-\phi}]\}. \tag{7}$$

Therefore, the extra fine-tuning condition obtains if the matter action is constrained to be scale invariant at the stationary point, or equivalently, if the vacuum is an infrared fixed point. In order that this condition be natural, it should emerge as a consequence of a symmetry of the effective theory. However, the no-go theorem in its general formulation seems to rule out this possibility [3]. This raises the question which is the focus of this letter: Does the action possess any helpful pseudo-symmetries?

Any symmetry transformation relevant to the cosmological constant problem necessarily involves the metric field. With this in mind, note that $S_m$ is invariant under the pseudo-symmetry transformation:

$$\begin{aligned} \phi &\longrightarrow \phi + a \\ g_{\mu\nu} &\longrightarrow e^{-2a} g_{\mu\nu} \\ \Delta_0 &\longrightarrow \Delta_0 - a\Delta_1 \\ \Delta_2 &\longrightarrow e^a \Delta_2. \end{aligned} \tag{8}$$

It is convenient to eliminate $\Delta_0$ by the equation of motion, Eq. (5), and reexpress the action in the form

$$S_m[\phi, g] = \int d^4x (\sqrt{g} e^{4\phi_o}) \{-e^{4(\phi-\phi_o)}[\tilde{d}_1(\phi - \phi_o - \frac{1}{4}) + \tilde{d}_2(e^{-(\phi-\phi_o)} - \frac{3}{4})]\}, \tag{9}$$

where $\tilde{d}_1 \equiv \Delta_1$ and $\tilde{d}_2 \equiv \Delta_2 e^{-\phi_o}$. This change of "basis" serves to illustrate that a shift in the coupling constants simply reparametrizes the dilaton VEV; $\tilde{d}_1$ and $\tilde{d}_2$ are, by construction, invariant under the transformation of Eq. (8), and so in the new basis, $S_m$ is manifestly invariant under the transformation

$$\begin{aligned} \phi &\longrightarrow \phi + a \\ g_{\mu\nu} &\longrightarrow e^{-2a} g_{\mu\nu} \\ \phi_o &\longrightarrow \phi_o + a. \end{aligned} \tag{10}$$

*A priori*, this transformation (or equivalently that of Eq. (8)) is not a pseudo-symmetry of the full theory, since the Einstein action, Eq. (4), is not an invariant. Moreover, the rescaling of the metric renders the vacuum energy $\phi_o$-dependent: $\langle T_\mu^\mu(\phi_o) \rangle = e^{4\phi_o}(\tilde{d}_1 - \tilde{d}_2)$. This is clearly problematic since the vacuum energy is a real measurable quantity in the presence of gravity. It is still possible to realize this invariance on the full theory; $S_m$ must be constrained to vanish at the stationary point ($\tilde{d}_1 = \tilde{d}_2$). Although the Einstein action is rescaled by the transformation, the Einstein equation, $R_{\mu\nu} = 0$, is clearly an invariant (flat spacetime looks the same at all scales). One might expect the overall normalization of the Einstein action (quantum corrections) to become relevant when higher dimensional operators like $\alpha R^2$ become important[4]. It is therefore gratifying that on general grounds this operator is suppressed relative to the Einstein term by a factor $q^2/M_p^2$, where $q \sim m_\phi$ is a characteristic momentum in the effective theory. What about other sectors of the effective theory? Clearly Eq. (10) is an invariance of the full effective theory only if all operators involving matter and gauge fields are scale invariant; that is, only the dilaton potential can break the scale invariance. This is easily accomplished by inserting appropriate powers of the dilaton field where necessary (for example, $-\sqrt{g}m\bar{\psi}\psi e^\phi$ is an invariant fermionic mass term). There is then a sense in which the cosmological constant is a natural parameter [9] in this toy effective theory, since in its absence, there is an enhanced pseudo-symmetry under the transformation of Eq. (10).

---

4 Note that this operator is present even if the gravitational field is not quantized [8].

Now we will show that the pseudo-symmetry, Eq. (10), implies a vanishing vacuum energy, independent of the specific form of the potential. Note that in the toy model, for finite values of $\phi_o$, $\langle T_\mu^\mu(\phi_o)\rangle$ is independent of $\phi_o$ if and only if it vanishes. This suggests that the vacuum energy might vanish by an equation of constraint. Consider the classic example of reparametrization invariance: electrodynamics. Gauge invariance renders the time component of the vector potential, $A_o$, arbitrary. So one can always choose $A_o = 0$. However, one must also *impose* the condition, $\delta S/\delta A_o = 0$ (Gauss's law), as an equation of constraint. In the present context, consider the following general argument with any number of scale anomalies, $N_a$. Pseudo-symmetry of $S_m$ can be expressed via the Ward identity:

$$0 = \delta g_{\mu\nu}\frac{\delta S_m}{\delta g_{\mu\nu}} + \delta\phi\frac{\delta S_m}{\delta\phi} + \sum_i \delta\Delta_i\frac{\delta S_m}{\delta\Delta_i}, \qquad (11)$$

where $i$ ranges from 0 to $N_a$. The $\Delta_i$ represent the undetermined dimensional parameters in the potential. The existence of a non-trivial equilibrium point implies

$$0 = \delta g_{\mu\nu}\frac{\delta S_m^o}{\delta g_{\mu\nu}} + \sum_i \delta\Delta_i\frac{\delta S_m^o}{\delta\Delta_i}, \qquad (12)$$

where the superscript signifies that the action is at the equilibrium point. The freedom in choice of $\phi_o$ can be expressed via the chain rule as

$$\frac{\delta S_m^o}{\delta\phi_o} = \sum_i \frac{\delta\Delta_i}{\delta\phi_o}\frac{\delta S_m^o}{\delta\Delta_i} = \frac{\delta S^o}{\delta\phi_o}. \qquad (13)$$

In the last step, use has been made of the fact that only the potential varies with $\phi_o$. The definition of the energy-momentum tensor can then be used to obtain

$$\langle T_\mu^\mu\rangle = \frac{1}{\sqrt{g}}\frac{\delta S[\phi_o, g]}{\delta\phi_o}. \qquad (14)$$

The assumed pseudo-symmetry implies that the effective theory is invariant under relabelings of $\phi_o$, and so the equation of constraint guaranteeing an invariant vacuum energy is simply $\langle T_\mu^\mu\rangle = 0$. A vanishing vacuum energy is therefore a natural consequence of the pseudo-symmetry, Eq. (10). Several general comments are in order. We must have $N_a > 0$ in order that an equilibrium point exist. However, it is easy to see that unless $N_a \geq 2$, the system is overconstrained and the potential vanishes. The toy matter action, subject to

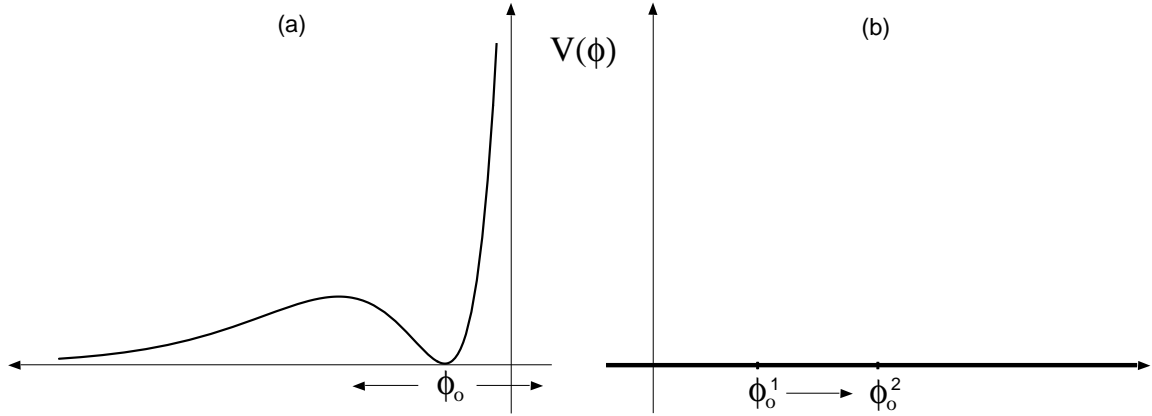

**Figure 1:** The dilaton potential. There are two ways in which the pseudo-symmetry of Eq. (10) can be realized. In (a), the potential is non-zero and has a global minimum where the potential vanishes (non-supersymmetric realization). In (b), the potential is flat (supersymmetric realization), and the pseudo-symmetry is implicit.

the equation of constraint, can be expressed in the form

$$S_m[\phi, g] = \int d^4x (\sqrt{g} e^{4\phi_o}) \{m^2 M^2 e^{4(\phi-\phi_o)}[1 - (\phi - \phi_o) - e^{-(\phi-\phi_o)}]\}, \qquad (15)$$

where $m^2 = m_\phi^2$ when $\phi_o = 0$. The potential is plotted in Figure 1a.

What has been gained by this analysis? The name of the game is to find a symmetry principle which implies a vanishing cosmological constant at macroscopic scales. This analysis has identified a pseudo-symmetry, Eq. (10), which appears to accomplish this task. The questions of interest are then: (i) Why should the macroscopic effective theory respect this invariance? and (ii) What is the origin of the dilaton? One possibility is explored below.

A generic property of string theory is the existence of a massless mode, the dilaton, which acts as the string coupling constant [10]. In superstring theories the graviton and the dilaton are in the same supermultiplet, and therefore the dilaton is constrained to be massless above the scale of supersymmetry breaking, $m_{SUSY}$. Furthermore, supersymmetry forbids a non-zero potential. There is thus a large vacuum degeneracy. These degenerate vacua are related by a classical scale invariance, which is an exact property of superstring theory [11]. However, there is also an *implicit* reparametrization invariance, since the inequivalent vacua are labeled by the dilaton VEV; that is, the theory also has a simple classical invariance corresponding to an arbitrary shift in the dilaton VEV. Hence, the combined tranformation, Eq. (10), is an exact classical pseudo-symmetry of superstrings[5].

---

5 Of course, here we assume a particular definition for the metric, as well as suitable scaling tranformation laws for the gauge and matter fields in the string-inspired effective theory.

Presumably quantum effects break supersymmetry and lift the vacuum degeneracy. Therefore, in the quantum theory, both the scale invariance and the invariance under shifts in the dilaton VEV are separately broken. However, since the combined pseudo-symmetry, Eq. (10), can be realized in the presence of a non-vanishing dilaton potential, in principle it could survive in the quantum theory. In other words one can think of the pseudo-symmetry as having a supersymmetric realization (Figure 1b) in which a pseudo-symmetry transformation relates inequivalent vacua, as well as a non-supersymmetric realization (Figure 1a) in which a pseudo-symmetry transformation relates equivalent vacua (as in the toy model discussed above).

This conjecture does not seem to violate any model-independent expectations. The dilaton is expected to have an exponential potential below $m_{SUSY}$ with certain general characteristics [12]. If the potential is perturbative in the string coupling constant, $g_s = e^\phi$, it should be an expansion in $g_s^2$ with model-dependent coefficients (determined by the mechanism that breaks supersymmetry). Non-perturbative effects would appear in powers of $1/g_s^2$. Moreover, the potential should vanish as $\phi \to -\infty$, which corresponds to the weak coupling limit (as in Figure 1a). Evidently, a potential with the necessary properties can arise only if string theory is strongly coupled, since leading order cannot stabilize the potential [13]. What about the matter and gauge couplings? Here the pseudo-symmetry has a non-trivial consequence. As noted above, the pseudo-symmetry allows the breaking of scale invariance only in the potential. It therefore requires that the dilaton potential be strongly coupled, while the rest of the effective theory appear weakly coupled (the scale invariance must continue to hold in the quantum theory for operators involving gauge and matter fields ). This consequence is in fact desirable, since a stable minimum implies strong coupling whereas all of the good features of string theory seem to rely on weak coupling [14].

There are independent phenomenological reasons for suspecting that a superstring dilaton is relevant to the cosmological constant problem. The fundamental bound on the dilaton mass is the observational bound on $\rho_{\rm VAC}$, as mentioned above. Roughly, one would expect [3] $\rho_{\rm VAC} \lesssim \rho_{\rm C}$ ($\rho_{\rm C}$ is the critical energy density). If the dynamics of the dilaton is responsible for the vanishing of the cosmological constant, it follows that $m_\phi \lesssim (\rho_{\rm C})^{\frac{1}{4}} \simeq 10^{-12} GeV$. Now consider the dilaton potential. On general grounds, one would expect $m_\phi^2 M^2 \simeq \mu^4$, where $\mu$ is the largest characteristic scale generated by the scale anomaly [4]. Consider a superstring dilaton. The dilaton and the graviton are now of common origin, and so $M = M_p$. Since the dilaton is expected to aquire a potential and therefore a mass at $m_{SUSY}$, it follows that $m_\phi^2 M_p^2 \simeq m_{SUSY}^4$. One further expects $m_{SUSY} \gtrsim 4\pi v$, where $v$ is the weak scale, and $4\pi v$ is a naive dimensional estimate of the electroweak symmetry breaking scale. This relation implies a theoretical lower bound on the mass of a superstring dilaton, $m_\phi \gtrsim 10^{-12} GeV$. Low-energy supersymmetry breaking, $m_{SUSY} \simeq 4\pi v$ ($m_\phi \simeq 10^{-12} GeV$), is then the only choice consistent with the cosmological bound on $m_\phi$. This choice has interesting cosmological implications. The vacuum energy is

$$\rho_{\rm VAC} \simeq \left(m_\phi\right)^4 \simeq \left(\frac{m_{SUSY}^2}{M_p}\right)^4 \simeq \left(10^{-12}\ GeV\right)^4 \sim \rho_{\rm C}. \qquad (16)$$

Theoretically this identity is interesting because it relates the hierarchy of scales to the large scale properties of the universe in a natural way. Moreover, cosmologies with $\rho_{\rm VAC} \sim \rho_C$ have many attractive features [15]. In particular, cold dark matter models with $\Omega_{\rm VAC} \simeq 0.8$ ($\Omega = 1$) are consistent with a wide variety of observations [16]. For example, it is well known that this is precisely what one needs in order to reconcile the recent measurement of the extragalactic distance scale [17] with estimated ages of galactic globular structures [18]. Recently, several independent groups [17] have extracted a Hubble constant of $H_o = 80 \pm 17\, km\, s^{-1}\, Mpc^{-1}$, which is in conflict with the age estimate of $16.5 \pm 2\, Gyr$ of Ref. 18, if one works in a flat universe with no cosmological constant. If one allows both matter and a positive cosmological constant ($\Omega_{\rm VAC} + \Omega_{\rm MATTER} = 1$), these measurements are consistent if $\Omega_{\rm VAC} \simeq 0.8$.

What about bounds that arise from short distance tests of Newtonian gravity? In order to extract an observational bound for the dilaton mass, one must estimate the coupling strength of the dilaton to matter. This is accomplished by considering variations in the strong coupling scale expected from string inspired dilaton-matter couplings [4,19]. The relevant experiments in the millimeter-centimeter range are of "Cavendish" type, which probe deviations from the inverse-square law. The authors of Ref. 19 have calculated a lower bound on the mass of a tree level string dilaton using the data of Ref. 20. At the $2\sigma$ level they find $m_\phi \gtrsim 1.6 \times 10^{-12} GeV$. Using a similar estimate for the coupling strength, the authors of Ref. 21 placed a lower bound on the mass of a superstring dilaton based on Eötvös-type experiments and satellite observations, $m_\phi \gtrsim 2 \times 10^{-13} GeV$. These bounds suggest that improved measurements in the millimeter range could discover the superstring dilaton.

There is, however, a generic cosmological problem endemic to light scalars with gravitational strength couplings. At an earlier epoch, one would expect the dilaton to be shifted from its minimum by finite-temperature effects, or by inflation. Coherent oscillations about the minimum could lead to an epoch in which the dilaton dominates the energy density of the universe [22,23,24]. Here one would expect $\Gamma_\phi \sim m_\phi^3/M_p^2$ which leads to $t_\phi \sim 10^{33} t_u$, where $t_u$ is the age of the universe, and so the dilaton is effectively stable. Hence the "misalignment" energy, acting as a form of dark matter, could overclose the universe in the present epoch, in blatant contradiction with observation. Although this problem renders the possibility of a light dilaton less attractive, the mechanism for a vanishing vacuum energy described above might act as a natural damping mechanism, since evidently the pseudo-symmetry constrains the dilaton to be at or very near its zero-curvature minimum at all scales. ( Presumably inflation would have to be driven by other fields.) Moreover, that these issues should be interrelated is appealing, since it is difficult to envision a solution of the cosmological constant problem, which is consistent with the effective theory paradigm, and yet does not involve at least one light scalar field with couplings of gravitational strength[6].

One frequently encounters the statement that understanding of why the cosmological constant is so small can emerge only from a consistent quantum theory of gravity. Athough possibly correct, it is not often appreciated just how radical this statement is. In fact, if the most conservative possibility is realized, a symmetry mechanism will be identified, involving as yet unobserved quanta, which operates at the scale of a mili-eV — a scale familiar to condensed matter physicists, and 31 orders of magnitude removed from the Planck scale. This conservative viewpoint rests on the spectacular success of local field theory. A general no-go theorem threatens the convervative point of view, and so should be subjected to extreme scrutiny. In this letter, it has been proposed that the no-go theorem might be avoided by assigning transformation properties to coupling constants as well as fields; i.e. via a pseudo-symmetry. In the context of a macroscopic effective theory with non-linearly realized scale invariance, an enhanced pseudo-symmetry has been identified in the absence of vacuum energy.

---

6 See also Ref. 1 and Ref. 23.

The superstring argument presented above is much in the spirit of a non-renormalization theorem [25]. The classical string-inspired effective action is invariant under a pseudo-symmetry transformation, which corresponds to a scale transformation accompanied by a shift in the dilaton VEV. We know that quantum corrections must break supersymmetry and lift the vacuum degeneracy. However, let us assume that the pseudo-symmetry survives in the quantum theory and use it to constrain the form of the low-energy effective action below the scale of supersymmetry breaking. The pseudo-symmetry constrains all matter and gauge couplings to be scale invariant (appear weakly coupled from point of view of string theory). The dilaton potential can have any number of terms, but it must have a global minimum where the potential vanishes (strongly coupled from string theory point of view). Hence, the pseudo-symmetry ensures that the vacuum energy remains zero after supersymmetry breaking.

One might worry that the light dilaton discussed above is ruled out by experiment. On the contrary, simple dimensional arguments suggest that if a superstring dilaton is relevant to the cosmological constant problem, then its range should fall precisely in the region of parameter space with the weakest observational bounds. This in turn implies a non-zero vacuum energy density of the order of magnitude required by "best-fit" cosmological models.

I am grateful for valuable conversations with A. Kapulkin, S.B. Liao, M. Mandelberg, C. Ordoñez, and especially B. Müller. I thank M. Strickland for help with the figure.